\documentclass[nofootinbib,aps]{revtex4}

\usepackage{amssymb,amstext,amsmath,amsthm}
\usepackage[dvips]{graphicx}
\usepackage{latexsym}
\usepackage{psfrag}
\usepackage{amsfonts}
\usepackage{bbm}

\newcommand{\Ref}[1]{(\ref{#1})}

\def\nn{\nonumber}

\newcommand{\eqa}{\begin{eqnarray}}
\newcommand{\neqa}{\end{eqnarray}}
\newcommand{\equ}{\begin{equation}}
\newcommand{\nequ}{\end{equation}}
\renewcommand{\nn}{\nonumber}

\newcommand{\be}{\begin{equation}}
\newcommand{\ee}{\end{equation}}
\newcommand{\bes}{\begin{eqnarray}}
\newcommand{\ees}{\end{eqnarray}}
\newcommand{\h}{\hspace{1mm}}

\newcommand{\N}{\mathbb{N}}

\newcommand{\R}{\mathbb{R}}
\DeclareMathOperator{\tr}{tr}

\def\arr{\rightarrow}

\newcommand{\V}{\mathbb{V}}
\newcommand{\T}{\mathcal{T}}
\newcommand{\VI}{\mathbb{V}_I}

\def\f{\frac}

\usepackage{bbm}
\def\C{{\mathbbm C}}

\newcommand{\SU}{\mathrm{SU}}
\newcommand{\SO}{\mathrm{SO}}

\def\dag{^\dagger}

\let\eps=\epsilon

\def\PP{{\cal P}}
\def\QQ{{\cal Q}}

\newcommand{\Spin}{\mathrm{Spin}}

\def\pp{\partial}
\def\vphi{\varphi}

\def\ka{\kappa}

\begin{document}

\title{{\large\bf 3d Spinfoam Quantum Gravity: \\Matter as a Phase of the Group Field Theory}}
\author{{\bf Winston J. Fairbairn}\footnote{winston.fairbairn@ens-lyon.fr} }
\author{{\bf Etera R. Livine}\footnote{etera.livine@ens-lyon.fr} }
\affiliation{\vspace{2mm}Laboratoire de Physique - ENS Lyon, CNRS-UMR 5672,  46 All\'ee d'Italie, 69364 Lyon, France}

\begin{abstract}
An effective field theory for matter coupled to three-dimensional quantum gravity was recently
derived in the context of spinfoam models \cite{effqg}. In this paper, we show how this relates
to group field theories and generalized matrix models.
In the first part, we realize that the effective field theory can be recasted as a matrix model
where couplings between matrices of different sizes can occur. In a second part, we provide a
family of classical solutions to the three-dimensional group field theory. By studying
perturbations around these solutions, we generate the dynamics of the effective field theory. We
identify a particular case which leads to the action of \cite{effqg} for a massive field living in
a flat non-commutative space-time. The most general solutions lead to field theories with
non-linear redefinitions of the momentum which we propose to interpret as living on curved
space-times. We conclude by discussing the possible extension to four-dimensional spinfoam models.
\end{abstract}

\maketitle



\section{Introduction: 2d and 3d Group Field Theories}


A large class of constrained or deformed topological BF field theories can be covariantly quantized
via the spinfoam technology \cite{SF}. In particular, spinfoam models provide viable candidates for
a covariant theory of quantum gravity in arbitrary spacetime dimensions
\cite{BC,BCd}. They can be regarded \cite{rr,SFrep} as implementing the physical scalar product in
loop quantum gravity by defining spacetime histories interpolating between the spin network quantum
states of the gravitational field.

Technically, they appear as two-complexes colored by statistical-like weights assigned to the
faces, edges and vertices of the dual two-skeleton of a chosen triangulation of spacetime. If the
theory is topological, the discretized theory is equivalent to the continuum theory since no local
degrees of freedom are present. If the theory under consideration does admit local degrees of
freedom, one needs to suppress the dependence on the triangulation. This procedure can be
implemented by computing a sum over triangulations in order to recover the infinite number of
degrees of freedom of the theories of interest. This is achieved by considering dual field theory
formulations named group field theories (GFT).

A $d$-dimensional GFT \cite{gftlaurent,dan_review} is an abstract, non-local field theory of
simplicial geometries living on ($d$ copies of) a group manifold. The theory is built such that its
Feynman evaluations reproduce the associated dual spinfoam amplitudes on the two-complexes given by
the Feynman diagrams of the GFT, consequently generalizing the matrix model (MM) technology to
higher dimensions. In fact, there is a crystal clear relationship between GFT's in two dimensions
and MM's. Let $G$ denote an arbitrary semi-simple, compact Lie group.

\bigskip

{\bf The $2d$ GFT} is a theory of complex, non-local fields $\vphi : G^{\times 2} \rightarrow \C$
satisfying a reality condition and a right invariance property under the diagonal $G$ action:
\be
\label{symmetry}
\vphi(g_1,g_2) = \overline{\vphi}(g_2,g_1) \hspace{3mm} \mbox{and} \hspace{3mm} \vphi(g_1 h,g_2 h) = \vphi(g_1,g_2), \hspace{2mm} \forall h \in \h G.
\ee
The dynamics of the theory are encoded in the following action
\be
S_{2d}[\vphi]=\f12 \int_{G^{\times 2}} dg_1dg_2\,\vphi(g_1,g_2)\vphi(g_2,g_1)
-\sum_n\f{\lambda_n}{n!}\int_{G^{\times n}} \prod_{i=1}^n dg_i
\vphi(g_1,g_2)\vphi(g_2,g_3)..\vphi(g_n,g_1),
\ee
where the $\lambda_n$'s are coupling constants and $dg_i$, $i=1,...,n$, denotes the normalized Haar measure on the $i^{th}$ copy of the compact group $G$. Because of the right invariance symmetry of the theory, we are free to fix the gauge and reformulate the theory in terms of a field $\phi$ on a single copy of $G$ defined as
$\vphi(g_1,g_2)\equiv\phi(g_1g_2^{-1})$. The $2d$ GFT action then reads:
\be
S_{2d}[\phi]=\f12\int_{G^{\times 2}} dg\,\phi(g)\phi(g^{-1}) -\sum_n\f{\lambda_n}{n!}\int_{G^{\times n}}\prod_{i=1}^n dg_i
\delta(g_1g_2..g_n)\phi(g_1)\phi(g_2)...\phi(g_n).
\ee
We can decompose the field $\phi$ along the unitary, irreducible representations $(\rho,
\mathbb{V}_{\rho})$ of $G$ using the Peter-Weyl theorem: $L^2(G) \simeq \bigoplus_{\rho}
\mathbb{V}_{\rho} \otimes \mathbb{V}_{\rho}^*$. We obtain the following expansion
\be
\phi(g)=\sum_{\rho} d_{\rho} \h \phi_{\rho \, ab} \h \rho(g)^{ab}.
\ee
Here, $d_{\rho} \in \N$ is the dimension of the representation $\rho : G \rightarrow Aut \h
(\mathbb{V}_{\rho})$, the indices $a,b = 1,..., d_{\rho}$ are matrix indices associated to the
matrix $\rho (g)$ representing the group element $g$, and $\phi_{\rho} \in \h \mathbb{V}_{\rho}
\otimes \mathbb{V}_{\rho}^* \simeq End \h (\mathbb{V}_{\rho})$ is the matrix Fourier coefficient of
the function $\phi$.

In other words, each $\phi_{\rho}$ is a rank $d_{\rho}=N$ matrix. More precisely, the symmetry
requirement of equation \eqref{symmetry} implies that the matrices are Hermitian
$\phi_{\rho}=(\phi_{\rho})\dag$, i.e., $\phi_{\rho} \in \mathfrak{u}(N)$. The $2d$ GFT action is
then re-expressed in terms of Hermitian matrix fields :
\be
S_{2d}[\phi_{\rho}]= \sum_{\rho} d_{\rho} \left[ \f12 \tr(\phi_{\rho}^2)-\sum_{n}\f{\lambda_n}{n!} \tr(\phi_{\rho}^n) \right].
\ee
This is the action of a tower of decoupled matrix models\footnotemark of all possible sizes
$N=d_{\rho} \in \N$. In the case of interest here, where $G=\SU(2)$, the unitary, irreducible representations
are labeled by spins $I\in\N/2$ and the sum is over half-integers.

The Feynman evaluation $I_{\mbox{{\tiny GFT}}}(\T^*)$ associated to a given diagram $\T^*$ equates
the topological invariant partition function $\mathcal{Z}_{\mbox{{\tiny BF}}}(\T)$ of
two-dimensional BF theory discretized on the triangulation $\T$ dual to the two-complex $\T^*$
defined by the GFT diagram :
\be
I_{\mbox{{\tiny GFT}}}(\T^*) = \mathcal{Z}_{\mbox{{\tiny BF}}}(\T) \equiv \sum_{\rho} d_{\rho}^{\chi(\T)},
\ee
where $\chi(\T)$ denotes the Euler characteristic of of the surface triangulated by $\T$, or
equivalently, of the diagram $\T^*$. It is explicitly given in terms of the number of vertices $V$,
edges $E$ and faces $F$ of the diagram $\T^*$ by $\chi(\T^*)=F-E+V$. This explains the presence of
this Euler characteristic in the Feynman evaluation. Indeed, each face corresponds to a closed
loop, i.e. a tracing leading to a $d_{\rho}$ factor. Similarly, each vertex carries a $d_{\rho}$
contribution since the dimension factorizes the whole action. Finally, each edge is associated to a
propagator which is trivially given by inverting the kinetic term of the action and yields
accordingly a $1/d_{\rho}$ contribution.

The sum over representations collapses to a single term if one considers BF theory on a surface
with boundaries supporting the canonical (spin network) states. These states select a particular
value of $\rho$ \cite{2dsf} which, from the MM perspective regarded as zero dimensional (scalar)
QCD, corresponds to the representation in which the quarks transform, i.e., to the number of colors
involved in the theory.

\footnotetext{
To avoid the trivial matrix model of size $1\times 1$ corresponding to the trivial $\rho_0$ mode, we can
require the field $\phi$ to have a vanishing integral over $G$: $\phi_0=\int \phi =0$.}

The same type of rationale remains true in higher dimensions. We now describe the three-dimensional case.

\bigskip

{\bf The $3d$ GFT} defines a manifold independent, covariant formulation of $3d$ quantum gravity.
Three-dimensional general relativity is a topological field theory and can accordingly be quantized
trough the spinfoam procedure. The resulting discretized path integral is called the Ponzano-Regge
(PR) model \cite{PR} and was actually the first quantum gravity model ever written. The dual GFT,
considered by Boulatov \cite{3dsf}, is defined on the Cartesian cube $G^{\times 3}$, $\vphi :
G^{\times 3} \rightarrow \C$, where $G=\Spin(\eta)$ with $\eta$ the diagonal form of a three
dimensional metric. For simplicity, we will consider Riemannian quantum gravity in the sequel, that
is, work in Euclidean signatures where $\SO(\eta) = \SO(3)$ and $\Spin(\eta) = \SU(2)$. To ensure
that the theory is torsion free, one requires a global right invariance of the field under diagonal
$G$ rotations
\be
\vphi(g_1h,g_2h,g_3h)=\vphi(g_1,g_2,g_3), \hspace{3mm} \forall h \in G.
\ee
Also, one requires the following reality condition :
\be
\vphi(g_1,g_2,g_3) = \overline{\vphi}(g_3,g_2,g_1).
\ee
The most general framework also requires some specific transformation properties of $\vphi$ under the permutation group
in order to generate all possible topologically inequivalent triangulations each with different weights (see e.g. \cite{dfkr} and references therein). We will neglect this aspect in the present work since it would unnecessarily complicate the derivation and eventually lead to the same results up to some numerical factors.

The dynamics are that of a non-local $\vphi^4$ theory :
\be
S_{3d}[\vphi]=\f12 \int_{G^{\times 3}} dg_1dg_2dg_3 \vphi(g_1,g_2,g_3) \vphi(g_3,g_2,g_1)
-\f\lambda{4!}\int_{G^{\times 6}}\prod_{i=1}^6dg_i\,
\vphi(g_1,g_2,g_3)\vphi(g_3,g_5,g_4)\vphi(g_4,g_2,g_6)\vphi(g_6,g_5,g_1).
\ee
The interaction term depicted above is a tetrahedron interaction chosen such that the GFT Feynman
diagrams generate (oriented) triangulations. More generally, we should in principle include
all closed spin network evaluations, which can then be interpreted as dual to arbitrary 3-cells.
For example, in \cite{laurent&david}, Freidel and Louapre include a ``pillow'' interaction term:
\be
S_{int}[\vphi]=\int_{G^{\times 6}} \prod_1^6dg_i\,
\vphi(g_1,g_2,g_3)\vphi(g_2,g_3,g_4)\vphi(g_4,g_5,g_6)\vphi(g_5,g_6,g_1),
\ee
which is the only other non-trivial quartic interaction term. Moreover they show that including it
renders the partition function Borel-sommable. From this perspective, including all possible
consistent interaction terms is natural in an effective QFT approach when studying the
renormalisation of the group field theory.

As for the two-dimensional case, we can expand the field $\vphi$ in term of the unitary,
irreducible representations\footnote{We will denote
$$
\VI = \C \{ \overset{I}{e}_a \mid a=-I,...,I \} = \C \{ \mid I, a > \}_{a} \simeq \C^{2I+1},
$$
for the complex vector space associated to the spin $I$ unitary, irreducible representation of
$\SU(2)$.} $(D^I, \mathbb{V}_I)$, $I \in \N/2$, of $G=\SU(2)$. For all triple of unitary,
irreducible representations $I, J, K$, let $\Psi^{K}_{\h IJ}: \VI \otimes \V_J \rightarrow \V_K$
and $\Phi^{IJ}_{\h K} : \V_K \rightarrow \VI \otimes \V_J$ denote the Clebsh-Gordan intertwining
operators\footnotemark. Using the right invariance property of the field, one obtains the following
decomposition
\be
\vphi(g_1,g_2,g_3) = \vphi_{I_1I_2I_3} \, . \left( \bigotimes_{i=1}^3 \sqrt{d_{I_i}} D^{I_i}(g_i) \right) . \, \iota^{\dagger I_1I_2I_3}.
\ee
\footnotetext{The Clebsh-Gordan coefficients well known from the quantum mechanics of
angular momentum are defined by the following evaluations
$$
\Phi^{IJ}_{\h K}(\overset{K}{e}_c) = \sum_{a,b}
\left( \begin{array}{cc} a & b \\
                         I & J \end{array} \right. \left| \begin{array}{c} K \\                                                                                                                            c  \end{array}\right)
\overset{I}{e}_a \otimes \overset{J}{e}_b, \hspace{3mm} \mbox{and} \hspace{3mm}
\Psi^{K}_{\h IJ}( \overset{I}{e}_a \otimes \overset{J}{e}_b) = \sum_{c}
\left( \begin{array}{c} c \\
                        K  \end{array} \right| \left. \begin{array}{cc}  I & J \\
                                                                         a & b \end{array} \right)
\overset{K}{e}_c $$.}
Here, the dots `.' stand for tensor index contraction. The normalized, three-valent intertwining operator
$\iota_{I_1I_2I_3} \in \h Hom_G(\mathbb{V}_{I_1} \otimes \mathbb{V}_{I_2} \otimes \mathbb{V}_{I_3}, \C)$ is related to the Clebsch-Gordan map $\Psi^{I_3}_{\h I_1I_2}$ by
the following evaluation called a $3j$ symbol
\be
\iota_{I_1I_2I_3}(\stackrel{I_1}{e_{a_1}} \otimes \stackrel{I_2}{e_{a_2}} \otimes \stackrel{I_3}{e_{a_3}}) =
\left( \begin{array}{lll}
I_1 & I_2 & I_3 \\
a_1 & a_2 & a_3
\end{array} \right) = e^{i \pi(I_1-I_2-I_3)} \h (\stackrel{I_3}{e_{a_3}} \h , \Psi^{I_3}_{\h I_1I_2} (\stackrel{I_1}{e_{a_1}} \otimes \stackrel{I_2}{e_{b_2}}) ) \h \in \R,
\ee
where $(,)$ is the scalar product associated to the bijective intertwiner $\epsilon_I :
\mathbb{V}_{I} \rightarrow \mathbb{V}_{I}^*$. The symbol $\iota^{\dagger I_1I_2I_3} \in \h Hom_G
(\C, \mathbb{V}_{I_1} \otimes \mathbb{V}_{I_2} \otimes \mathbb{V}_{I_3})$ denotes the associated
adjoint operator. The ``tensor'' fields $\vphi_{I_1I_2I_3}$ are given in terms of the Fourier modes
$\tilde{\vphi}^{I_1I_2I_3}$ of the Peter-Weyl decomposition of the GFT field by $\vphi_{I_1I_2I_3}
= \tilde{\vphi}^{I_1I_2I_3} \iota_{I_1I_2I_3} \prod_{i=1}^3 \sqrt{d_{I_i}}$. The Boulatov theory
can then be understood as a generalized matrix model based on $3$-tensors instead of matrices. Each
tensor field corresponds to an elementary two-simplex whose propagation builds simplicial
three-geometries. The action reads
\be
S_{3d}[\vphi_{I_1I_2I_3}] = \sum_{\{ I \}} \left[ \f12 \mid \vphi_{I_1I_2I_3} \mid^2 - \frac{\lambda}{4!} \vphi_{I_1I_2I_3} \vphi_{I_3I_5I_4} \vphi_{I_4I_2I_6} \vphi_{I_6I_5I_1}
\left\lbrace \begin{array}{lll}
I_{1} & I_{2} & I_{3} \\
I_{4} & I_{5} & I_{6}
\end{array} \right\rbrace \right],
\ee
where the summation symbol is a sum over all spins $I$ (and implicitly all magnetic numbers $a$)
appearing at the right hand side of the sum and
\be
\left\lbrace \begin{array}{lll}
I_{1} & I_{2} & I_{3} \\
I_{4} & I_{5} & I_{6}
\end{array} \right\rbrace = \left( \begin{array}{lll}
a_1 & a_2 & a_3 \\
I_1 & I_2 & I_3
\end{array} \right)
\left( \begin{array}{lll}
I_1 & a_5 & I_6 \\
a_1 & I_5 & a_6
\end{array} \right)
\left( \begin{array}{lll}
I_4 & I_2 & a_6 \\
a_4 & a_2 & I_6
\end{array} \right)
\left( \begin{array}{lll}
a_4 & I_5 & I_3 \\
I_4 & a_5 & a_3
\end{array} \right),
\ee
is the $6j$ symbol in a particular orientation configuration.

Once again, the GFT is built such that its Feynman diagram evaluations generate the path integral
of three-dimensional gravity discretized on the triangulations dual to the two-complexes defined by
the diagrams. Indeed, if $\T^*$ denotes a Feynman diagram of Boulatov's GFT, we have the following
equality
\be
I_{\mbox{{\tiny GFT}}}(\T^*) = \mathcal{Z}_{\mbox{{\tiny PR}}}(\T) \equiv \sum_{\{I_e\}} \prod_{e} d_{I_e} \prod_t \left\lbrace \begin{array}{lll}
I_{1t} & I_{2t} & I_{3t} \\
I_{4t} & I_{5t} & I_{6t}
\end{array} \right\rbrace ,
\ee
where the products are respectively on the edges $e$ and tetrahedra $t$ of the triangulation $\T$,
dual to the cellular complex $\T^*$, on which the Ponzano-Regge model is defined.

To summarize, GFT's appear as extremely powerful tools to generate simplicial manifolds of
arbitrary topologies and dimensions. Their use is furthermore highlighted in the study of
non-topological theories where one needs to recover the infinite number of degrees of freedom from
lattice discretized models. That said, a clear physical interpretation of the GFT is still lacking
despite some progress \cite{gftlaurent}. What is the physical content of the fields? What is the
meaning of the coupling constant? What are the classical symmetries? What are the non-perturbative
properties of these theories? These issues remain open. In the following, we reconsider $2d$ GFTs
as non-abstract theories and introduce a physical field theory sharing many GFT properties but
whose physical interpretation is perfectly controlled.

\bigskip

{\bf $2d$ spinfoams embedded into $3d$ spinfoams : effective field theory}

The coupling of point particles to the PR model \cite{pr1} has unraveled an intriguing relationship
between the Feynman diagrams of quantum field theory (QFT) and spinfoam models: Feynman diagrams
have been shown to yield natural quantum gravity observables \cite{aristide, barrett}. Moreover,
the expectation values of such observables, attached to graphs embedded in the spinfoam,
can be re-expressed as deformed Feynman amplitudes issued from a QFT on a flat, non-commutative
space-time. This leads to the notion of an effective field theory (EFT) describing the dynamics of
matter once that the quantum gravity fluctuations have been integrated out \cite{effqg}. The
non-commutativity in space-time coordinates is encoded in a well-defined $\star$-product (which is
not of Moyal type) whose detailed construction can be found in the original work \cite{pr3,effqg}.
The key point of that analysis is that although space-time remains flat, the momentum space
becomes a curved manifold $G$ isomorphic, as a manifold, to $\Spin(\eta)$. In the limit where the
Newton constant $G_N$ is sent to zero, the star-product becomes commutative, momentum space becomes
flat and one recovers ordinary QFT. The crucial point, that we will develop in the sequel, is the
fact that such Feynman diagrams are in fact two-dimensional spinfoams. Accordingly, we are led to
the study of spinfoam models embedded into higher dimensional quantum geometrical backgrounds. From
now on, we will concentrate on the Riemannian $G=\SU(2)$ case. We expect that the core of our
results will translate to the non-compact case.

Consider the generalized two-dimensional GFT defined by the (momentum space) action
\be
\label{gGFT}
S[\phi]=\f12 \int_{G} dg \, \phi(g) \h \mathcal{K}(g) \h \phi(g^{-1})
-\sum_n\f{\lambda_n}{n!}\int_{G^{\times n}}\prod_{i=1}^n dg_i \delta(g_1g_2...g_n) \prod_{i=1}^n \phi(g_i).
\ee
Here the group variables $g$ are now considered as deformed momenta of (scalar) particles
propagating in $3d$ quantum gravity, in contrast with the non-embedded, abstract $2d$ GFT. We will
assume that the usual symmetry requirements of the GFT on the scalar field $\phi$ have been
relaxed. The group function $\mathcal{K}$ denotes the kinetic kernel in momentum space, which is
assumed to obey an $Ad(G)$-invariance, that is, to be central. Finally, the $\lambda_n$'s are the
coupling constants for the $n$-vertex interactions of the scalar field and $\delta(g_1..g_n)$
describes the momentum conservation at each vertex.

Let $\Gamma \subset M$ denote a Feynman diagram of the theory. It can be embedded into a
triangulated two-surface $(\Sigma,\Delta)$, which in turn can be embedded into a $3d$ triangulation
$\T$ of the spacetime manifold $M$.
If the triangulation $\T$ is homeomorphic to the three-sphere, the Feynman amplitude $I(\Gamma)$
associated to the diagram $\Gamma$ can be written as a three-dimensional state sum model
$\mathcal{Z}(\Gamma,\T)$ via a topological duality transformation \cite{pr3,effqg} and integration
over the group (momentum) variables:
\be
I(\Gamma) = \mathcal{Z}(\Gamma,\T) \equiv \sum_{\{I_e\}} \prod_{e \notin \Gamma} d_{I_e} \prod_{e
\in \Gamma} \mathcal{P}(I_e) \prod_{t} \left\lbrace \begin{array}{lll}
I_{1t} & I_{2t} & I_{3t} \\
I_{4t} & I_{5t} & I_{6t}
\end{array} \right\rbrace ,
\ee
where $\mathcal{P}(I)$ is the Fourier mode appearing in the Peter-Weyl decomposition of the
propagator $\mathcal{P} \equiv i \mathcal{K}^{-1} = \sum_I \mathcal{P}(I) \chi^I$ in
terms of characters $\chi^I(g) = \tr D^I (g)$. It is a function of the length quantum numbers of
the three-dimensional quantum geometry theory. This is due to the fact that the two-dimensional
spinfoam is embedded into a (triangulated) three-dimensional spacetime. More precisely, the
discretization procedure of $3d$ gravity leading to the Ponzano-Regge model assigns a physical
length vector $l_e^a = \int_{e } e^a \h \in \R^3$, $a=1,2,3$, to each one-simplex $e$ of the chosen
triangulation $\T$, where $e^a = e^a_{\mu} dx^{\mu}$ is the dual co-frame expressed into a
particular local basis $\{ d x^{\mu} \}_{\mu}$ of the co-tangent space. In the quantum framework,
the lengths are quantized and take only discrete values (because we are working in the Riemannian
framework thus with a compact gauge group) encoded in the representation labels $I_e$ assigned to
the one-simplices $e$ of the triangulation. Accordingly, the mode $\mathcal{P}(I_e)$ can be
understood as the propagator assigned to the edge $e$ of the diagram in (quantum) position space;
it depends only on the length separating its boundary vertices $t(e)$, $s(e)$. It is the signature
of the presence of a particle, i.e., of the obstruction to the flatness of the connection along the
corresponding edge.

Let us now make the description more concrete and consider specific choices of propagators.

\begin{itemize}
\item $\mathcal{P}(g) = 1$. This leads to the ordinary $2d$ GFT action (provided the fields obey the appropriate symmetry
conditions) regarded from a three-dimensional perspective. The corresponding spinfoam amplitude is
that of the PR model with insertion of an observable fixing the representations label to zero along
the edges of the Feynman diagrams : $\mathcal{P}(I) = \delta_{I,0}$.

Since the spins are interpreted as length quantum numbers, we can readily see that the $2d$ GFT
forces the lengths of the one-simplices supporting the edges of the Feynman diagram to vanish. In
other words, the embedding of the diagrams of the MM into the (simplicial) spacetime manifold is
degenerate (non-injective). It defines an immersion where all the points of the surface are mapped
onto a point. This is due to the fact that the GFT propagator along an edge $e$ is given by a delta
function on position space: $G(x_{t(e)} - x_{s(e)}) = \delta (x_{t(e)} - x_{s(e)})$
in the abelian no-gravity limit. In the $G_N \neq 0$ case, the absence of resolution beyond
the Planck length in non-commutative spacetime implies the replacement of the delta function by its
deformed analogue which is also concentrated on zero lengths but has non-zero width (it is the
first Bessel function \cite{pr3}). It is interesting to remark that the obtained spinfoam amplitude
is in fact a gauged fixed Ponzano-Regge partition function \cite{pr1}. The gauge fixing procedure,
i.e., the killing of the sum over representations, occurs along the $2d$ GFT diagram $\Gamma$. Note
however that the GFT can generate the maximal trees \cite{pr1} usually used to gauge fix the gauge
symmetries in the Ponzano-Regge model, only at the classical, or tree level.

\item $\mathcal{P}(g) = \delta(g)$. This choice leads to a theory free of matter: there are no particles traveling in the diagram, i.e., there are no topological defects in space-time. As remarked in \cite{kirillGFT}, modifying the $2d$ GFT as to include such a propagation term leads to the exact Ponzano-Regge amplitude.

\item $\mathcal{P}(g) = i(P^2(g) - M^2 - i\epsilon)^{-1}$, where $\epsilon\arr 0^{+}$ is a regulator. Here, the momentum $P:\,G \rightarrow
\mathfrak{g}$ is defined through the projection of the group element $g$ on the basis of Pauli
matrices $\{\sigma^a\}_a$ spanning the tangent space $\mathfrak{su}(2)$ :
\be
P^a(g) :=  \f{\kappa}{2i} \tr(g \sigma^a),
\ee
with $a=1,2,3$ and $\ka$ is the Planck mass related to the Newton constant by $\ka = (4 \pi
G_N)^{-1}$. The renormalized mass $M=\kappa
\sin
\theta$, where $\theta = \frac{m}{\kappa}$ is expressed in term of the deficit angle $\theta$ of the conical singularity created
by the particle of bare mass $m$. It takes into account the gravitational feed-back. The
expectation value of the associated PR observable

\be
\mathcal{P}(I) \equiv \mathcal{P}_{\theta} (I) = \frac{i}{2\kappa^2} \frac{e^{-id_I(\theta - i \epsilon)}}{\cos \theta},
\ee
corresponds to the PR amplitude with the insertion of Feynman propagators along the particles worldlines.

\end{itemize}

It is interesting to see that the $2d$ GFT considered as a physical field theory (not just as a computational tool) whose Feynman diagrams are embedded into spacetime can generate three-dimensional Ponzano-Regge geometries. It suggests a tight link between two-dimensional and three-dimensional spinfoam models. This paper is devoted to the detailled study of this relationship.

\bigskip
The plan of the paper is the following. Having investigated the GFT structure of the EFT in the introduction,
we show that it can be reformulated as a MM in section II. The question of the effect of the non-trivial propagation term involving a squared momentum is analyzed. As we will see, it will result in a richer structure where couplings between matrices of different sizes can occur. We will describe in details the Feynman rules of this new theory and give an interpretation of the variables involved in terms of three-dimensional quantum geometry.

In section III, we discuss the derivation of the EFT from Boulatov's GFT. This issue has been tackled in
\cite{gftwithparticles}, \cite{kirillGFT} where the authors generalize the Boulatov GFT to include particles. Here, we will give a simpler answer, showing that the effective QFT is a particular phase of the $3d$ GFT. We will first define a particular dimensional reduction yielding the two-dimensional GFT from its $3d$ counterpart, by selecting a particular $2d$ phase of $3d$ quantum gravity. We will then exhibit a one-parameter family of solutions to Boulatov's field equations corresponding to non-trivial geometrical backgrounds. The effective QFT will then be understood as describing (surface-like) perturbations
around these particular instantons of the Boulatov model. This fact highlights the interpretation of matter as excited states of geometry in three space-time dimensions. In fact, in the most general case, we will see that the obtained perturbed action is that of the EFT corrected by a non-linear redefinition of the momentum in terms of higher degree momenta. We will interpret these corrections as the signature of a non-trivial background corresponding to a spacetime metric which is not flat.

Section $IV$ is a four dimensional outlook where we sketch how to apply the same techniques for the $4d$ GFT. We conjecture that we will be led to a field theory of string-like excitations.

Finally, the Appendix contains a seek of the classical solutions to the EFT and an explicit calculation of the general three-dimensional perturbations of Boulatov's GFT which may be needed in generalizing our work to the fermionic case.

\section{3d Non-Commutative Field Theory as a Matrix Model}

The fact that the effective field theory (EFT) in momentum space defined by the action \eqref{gGFT}
with kinetic kernel $\mathcal{K}(g)=P^2(g) - M^2$ is defined in terms of fields living on a group
manifold turns it into a two-dimensional GFT. We thus expect a reformulation as a MM. We now
explore in detail this interesting analogy.

The expression of the EFT as a MM goes through the recasting of the action in terms of
representation labels.
We first use the group structure of the momentum space manifold to develop the field $\phi$ in Fourier
modes. We then consider the matrices $\phi_{Iab}$ as the dynamical fields of the EFT. Note that, a
priori, these matrices are not Hermitian, unless we impose the reality condition
$\phi(g^{-1})=\overline{\phi}(g)$.
The action of the EFT then reads:
\be
\label{repeff}
S_{eff}[\phi_I]=\f12 \sum_{I,J} \phi_I \h \mathcal{K}^{IJ} \h \phi_J - \sum_{n,J} \f{\lambda_n}{n!} d_J \tr(\phi_J^n),
\ee
where we have used the notation $\phi_I \h \mathcal{K}^{IJ} \h \phi_J := \phi_{I ab} \h
\mathcal{K}^{IJ \, abcd} \h \phi_{J cd}$ and the kinetic kernel $\mathcal{K}^{IJ} \, \in Aut \,
(\V_I \otimes \V_J)$ is given by
\be
\mathcal{K}^{IJ} = d_I d_J \int_G dg \h \mathcal{K}(g) \h D^I(g) \otimes D^J(g^{-1}),
\ee
with $\mathcal{K}(g) = P^2(g) - M^2$.
To display the propagator $\mathcal{P}$, we next invert the
kinetic part of the action. Inserting matrix indices, we obtain the following propagation term
\be
\mathcal{K}^{-1IJ}_{\;\;ab \, cd} = \int_G dg \h \mathcal{K}(g)^{-1} \h D^{I}(g)_{ba} D^{J}(g^{-1})_{dc} ,
\ee
such that $\mathcal{K}^{IJ \, ab \, cd} \mathcal{K}^{-1JK}_{\;\; cd \, ef} = \delta^{IK} \delta^{a}_e \delta^b_f$.

We can now express the group function $\mathcal{K}^{-1}$ in the basis of the unitary, irreducible representations of $G$.
To this aim, we use the fact that $P^2(g) \equiv P^2(\alpha) = \kappa^2 \sin^2(\alpha)$ is $Ad(G)$-invariant and has also a central inverse
\footnote{Here, we are using the diffeomorphism mapping $\SU(2)$ onto the unit three-sphere $S^3$ in
quaternion space $\mathbb{H}$, to parametrize the group manifold through a radius $\alpha \in [0,
2\pi]$ and a unit vector $n \in S^2$. Accordingly, any group element $g$ is parametrized by
$g(\alpha,n)=\cos\alpha \h 1\!\!1 + i \sin \alpha \h n_a \sigma^a$, and the squared momentum yields
$P^2(g)=\kappa^2 \sin^2\alpha$.}. The (regulated) inverse kernel $\mathcal{K}^{-1}(\alpha) =
(\kappa^2(\sin^2\alpha - \sin^2\theta - i \epsilon))^{-1}$ can thus be expanded in terms of
characters
\be
\mathcal{K}^{-1}(g) = (P^2(g)-M^2 - i\epsilon)^{-1} = -i \sum_I \mathcal{P}_{\theta}(I) \chi^I (g),
\ee
with the kernel $\mathcal{P}_{\theta}(I)$  defined in the previous section.

Using the integration formula of the tensor product of three representations
\be
\int_G dg \h D^I(g) \otimes D^J(g) \otimes D^K(g^{-1}) = \frac{1}{d_K} \Phi^{IJ}_{\h K} \Psi^K_{\;\; IJ},
\ee
we obtain the propagator in representation space
\be
\label{propagator}
\mathcal{P}^{IJ} \equiv \h i \mathcal{K}^{-1IJ} = \sum_K \frac{\mathcal{P}_{\theta}(K)}{d_J} \Phi^{KI}_{\h J} \h \Psi^J_{\;\; KI}.
\ee
The Feynmanology of the EFT casted as a generalized MM follows immediately and is depicted in FIG. $1$.
\begin{figure}[t]
\begin{center}
\psfrag{i}{$I$}
\psfrag{j}{$J$}
\psfrag{k}{$K$}
\psfrag{P}{$ \h \equiv \h \sum_{K,e} \frac{\mathcal{P}_{\theta}(K)}{d_J}
\left( \begin{array}{cc} e & b \\
                         K & I \end{array} \right. \left| \begin{array}{c} J \\                                                                                                                            c  \end{array}\right)
\left( \begin{array}{c} d \\
                        J  \end{array} \right| \left. \begin{array}{cc}  K & I \\
                                                                         e & a \end{array} \right)
$}
\psfrag{V}{$ \h \equiv \h \delta^{IJ} \delta^{JK} \delta^e_a \delta^b_c \delta^d_f$}
\includegraphics{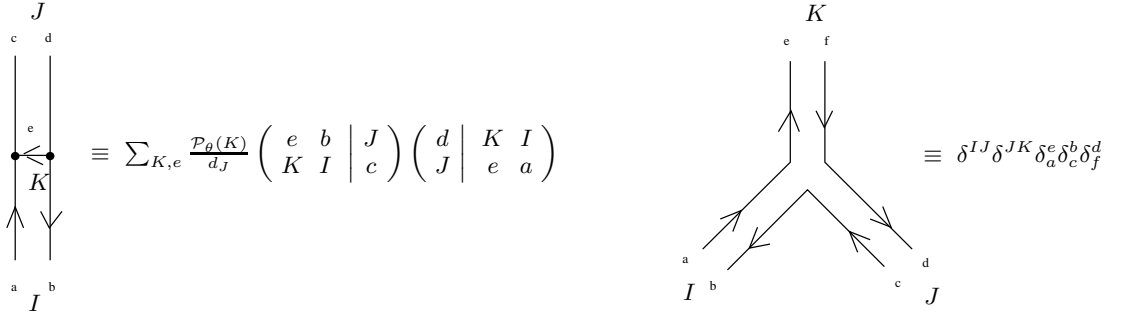}
\caption{Feynman rules for the EFT in representation space for the $n=3$ (trivalent interaction) case.
.}
\end{center}
\end{figure}
One can verify that setting $\mathcal{P}_{\theta}(K)=\delta_{K,0}$ in \eqref{propagator} leads to the correct trivial propagation of the GFT encountered previously, that is, $\mathcal{P}^{IJ}=\delta^{IJ}/d_J$.

We close this section with a set of remarks. The first remark is that the diagrams generated by the
EFT have a fat structure, namely the propagators are built out of two lines, each carrying a matrix
index. This property is shared by any theory whose dynamical fields have a matrix structure, like
for instance non-abelian gauge theories where the double index structure of the gauge field
(carrying e.g. color indices) can be geometrically interpreted as the composite nature of the gauge
particles \cite{Nexpansion}. Indeed, if the theory contains fermionic fields, they will carry only
one color index, as they live in color space, which implies that the gauge field can be
geometrically, and only in that sense, regarded as composed by a fermion and an anti-fermion. As a
result, the Feynmanology of the EFT generates two-dimensional piecewise-linear manifolds where the
two lines of the propagators correspond to the gluing of two elementary two-cells each defined by a
Feynman loop. In general, the MM's or $2d$ GFT's generate two-complexes dual to triangulations of
surfaces. The same is true here. However, if we restrict ourselves to the diagrams containing loops
of order three (going through three vertices), the theory generates two dimensional triangulations
which, when embedded in a three dimensional triangulation, are related via a duality transformation
to the insertion of particles along the one-simplices (propagators) of the $3d$ triangulation in
the Ponzano-Regge model.

Second, let us stress an important deviation from standard MM's which is the non-triviality of the
propagator (see FIG. $1$), taking into account the matter propagation. This aspect is also present
in string theory where the presence of matter fields (the embedding of the worldsheet in target
space) modifies the propagation properties of the dual MM formulation. The immediate consequence is
the dynamical coupling of different rank matrices. More precisely, the Feynmanology teaches us that
the matrices evolve dynamically into matrices of different sizes. This non-trivial fact has no
gauge field theory formulation counterpart. However, to parallel the geometrical composite
interpretation discussed above, we see that the matrix field of the EFT can be regarded as composed
by interacting `fermions' belonging to a theory where the number of colors changes dynamically
through interactions, like the number of spin states in ordinary QFT.

Finally, we can give a geometrical interpretation of the `color' labels appearing in the EFT, that
is, relate the surface theory generated by the EFT to three-dimensional quantum gravity. Here, the
size of the matrices is related to the length quantum numbers of the three-dimensional quantum geometry
theory because the EFT carries information about its embedding into a (triangulated) three-dimensional
spacetime. We have seen that the physical quantum length numbers are
given by the representation labels assigned to the one-simplices $e$ of the triangulation. At the classical level,
these physical length vectors are measured with respect to a particular frame. This frame is usually
chosen to be associated to a given trivialization of the spin bundle over the spacetime spin
manifold by a global section, piecewise constant in each tetrahedron of the triangulation. Let us
consider a such tetrahedron and focus on one of its boundary triangles assumed to support a
(three-valent) Feynman loop of the EFT. The loop is built out of three propagators each carrying a
spin on its end points. By virtue of the Feynman rules (FIG. 1), the representation labels before
and after a vertex are constrained to match. Accordingly, the loop carries three spin, say $I_1$,
$I_2$ and $I_3$ each sitting on one of the summits $123$ of the considered triangle, taken in
cyclic order. Regarding the geometrical interpretation discussed above, it appears clearly that the
three spins encode the quantum distance to the origin of the chosen frame. Accordingly, the
representation labels $I_{ij} = I_j - I_i$ assigned to the edges $ij$, $i,j = 1,2,3$, measure the
quantum lengths of the three boundary edges.
If we choose the reference frame appropriately, namely such that its origin coincides with one of
the summits, say the vertex $1$, and its three axis are along the three segments emerging from $1$,
the corresponding length accordingly reads $I_1=0$ and the quantum number $I_2 \equiv I$ measures
the length of the segment $12$. Concentrating on the associated propagator, we see that the fat
structure collapses
\bes
\mathcal{P}^{I0}_{ab\,00} &=& \sum_{K,e} \mathcal{P}_{\theta}(K) \h \delta_{KI} \h \epsilon_{Ibe} \epsilon_{Iea} \h  \\
                    &=&  \mathcal{P}_{\theta}(I) \h \delta_{ab},
\ees
where $\epsilon_{Iab}$ denote the matrix elements of the dual pairing intertwiner $\epsilon_I$ and
the delta function is on the vector space $\V_I$. Switching back to the
spinfoam perspective, we can readily see that we have recovered the propagator appearing in the PR
model as a sole function of the quantum length number $I$ associated to the corresponding
one-simplex.
This is the three-dimensional geometrical picture of the surface theory generated by
the EFT; the size of the matrix fields $\phi_{I_i}$ and $\phi_{I_j}$ living at the endpoints of the
propagator $ij$ encode the physical length of the one-simplex supporting the propagator.
The non-triviality of the EFT propagation does not constrain the rank of the two matrices to match
which would lead to one-simplices of zero physical length, as discussed in the introduction.
In this sense, the effective action recasted in terms of representation labels \eqref{repeff}
can be understood  as a QFT living on a discrete, quantum geometry background. The representation indices
$I$ are the counterparts of the position vectors $x$ in ordinary QFT.

We have just discussed the fact that the Feynman diagrams of the EFT are two-dimensional spinfoams whose
evaluations, once embedded into spacetime, yield three-dimensional spinfoam amplitudes of particles
propagating in the $3d$ PR geometry.  Now, we know that the PR partition function can be related to
the Feynman integrals of a $3d$ GFT. This leads to the following natural question: is it
possible to relate the EFT to a three-dimensional GFT? As we explain below, the EFT, and thus
matter excitations, can be understood as a $2d$ phase of Boulatov's $3d$ GFT.

\section{Matter as perturbations around classical solutions to the 3d GFT}

We now show how to obtain the EFT dynamics as perturbations around a non-trivial geometrical
background defined by a classical solution to the $3d$ GFT. As we are about to see, these
perturbations are of a special type, namely they are surface-like excitations. We call these types
of perturbations {\itshape two-dimensional phases} of Boulatov's theory.

We start by showing how to reduce the Ponzano-Regge model to its two-dimensional sub-sector. We
define the $2d$ phase of the Boulatov GFT by restricting the set of fields to those depending only
on two of their arguments which in turn are functions of a unique group element by virtue of gauge
invariance:
\be
\vphi(g_1,g_2,g_3)\,:=\, \psi(g_1,g_3) = \psi(g_1 g_3^{-1}).
\label{3dto2d}
\ee
Then it is easy to check that:
$$
S_{3d}[\vphi=\psi]
\,=\,
S_{2d}[\psi],
$$
because we are considering normalized Haar measures yielding a unit volume for the compact group
$G$. If we only consider the tetrahedron interaction term in the $3d$ GFT, we will obtain $\psi^3$
and $\psi^4$ interaction terms for the $2d$ phase. If we want to obtain interaction terms with
higher powers of the field $\psi$, we can include higher order terms in the $3d$ GFT.

From the point of view of the Fourier transform $\vphi^{I_1I_2I_3}$, the restriction $\vphi=\psi$
implies setting $I_2=0$ and thus reducing the 3-tensors to (square) matrices: the triangles of the
$3d$ GFT collapse to double lines since we have forced one of the boundary segments to have zero
length. Accordingly, we only generate $2d$ cellular decomposition and not $3d$ triangulations anymore;
the ansatz \Ref{3dto2d} defines the two-dimensional phase of the Boulatov group field theory.

Next, we introduce the `instantons' \footnotemark \footnotetext{Let us nevertheless point out that the
evaluation of the action on these solutions diverges because of a factor $\delta(1\!\!1)$.} around
which we are willing to study classical perturbations. Consider
the one parameter family of fields
\be
\label{solution}
\vphi_f(g_1,g_2,g_3)\,=\,\sqrt{\frac{3!}{\lambda}} \int_{G} dh\, \delta(g_1h) f (g_2h) \delta(g_3h),
\ee
which is parametrized by the function $f : \SO(3) \rightarrow \R$, an arbitrary function of
$L^2(\SO(3))$. The reality of $f$ is imposed by the reality condition constraining the fields of
the $3d$ GFT. The choice of $G=\SO(3)$ is to impose a purely even Fourier decomposition to avoid
non-analyticity issues. We also require $f$ to satisfy the following normalisation condition
\be
\int_G dg f^2 = 1.
\ee
An example of such a function is provided by the character $\chi^I$, $I \in \N$. This field
provides a whole family of solutions to the field equations of the Boulatov group field theory:
\be
\vphi(g_3,g_2,g_1)\,=\,
\f\lambda{3!}\int dg_4dg_5dg_6\,
\vphi(g_3,g_5,g_4)\vphi(g_4,g_2,g_6)\vphi(g_6,g_5,g_1).
\ee
Accordingly, $\vphi_f$ defines a non-trivial, three-dimensional background geometry. Even if the
link between GR and the GFT at the classical level is obscure, we make a first step toward
clarifying this issue in this paper and parallel the procedure of studying perturbations of
Einstein's theory around a given classical solution. Thus, we now question the classical GFT about
its behavior in the neighborhood of the solution $\phi_f$, i.e. study the first order perturbations
around this `instantonic' solution:
\be
S^{(f)}_{3d}[\vphi]\,\equiv\, \delta S_{3d} [\vphi] \, = \, S_{3d}\left[\, \vphi_f + \vphi\right]
-S_{3d}\left[\, \vphi_f\right].
\label{Sf}
\ee
We obtain the usual quadratic term in $\vphi^2$
corrected with $\vphi^2\vphi_f^2$ terms and the usual quartic interaction in $\vphi^4$ plus a new
$\vphi^3\vphi_f$ cubic term \cite{thesis} (see the Appendix for a detailed computation). For the
present analysis, we are more particularly interested in
{\it 2d perturbations}. To this aim, we expand
$$
\vphi(g_1,g_2,g_3)=\vphi_f(g_1,g_2,g_3)+\psi(g_1,g_3 ).
$$
This leads us to the following key result
\bes
S^{(f)}_{3d}[\psi]\,&=&\, \f1{\kappa^2}\int_G dg \h \psi(g) \, \mathcal{K}_f(g) \, \psi(g^{-1}) \\ \nn
&& -\f\mu{3!} \int_{G^{\times 3}} \prod_{i=1}^3
dg_i\,\delta(g_1...g_3) \prod_{i=1}^3 \psi(g_i) -\f\lambda{4!} \int_{G^{\times 4}} \prod_{i=1}^4
dg_i\,\delta(g_1...g_4) \prod_{i=1}^4 \psi(g_i).
\ees
Here, the kinetic term is given by
\be
\forall g \in G, \;\;\; \mathcal{K}_f(g) = \frac{\kappa^2}{2} \left[ 1 -2 \left( \int_{G} d h f(h) \right)^2 -1 \left(\int_G dh\,
f(h)f(hg)\right) \right],
\ee
where the quadratic contribution in $(\int f)^2$ comes from quartic interactions of the type $\int
\vphi_f\vphi_f\psi\psi$ while the convolution term in $f \circ f$ comes from $\int
\vphi_f\psi\vphi_f\psi$ contributions.
From here on, we require that $f$ is central\footnotemark, i.e $Ad(G)$-invariant. This is to ensure
that the kinetic term $\mathcal{K}(g)$ be constant on the conjugacy classes and thus depend only on the norm of the
momentum (and not its direction).
\footnotetext{Note however that there are no technical obstructions in relaxing the conjugation invariance
requirement. The theory would be non-isotropic, but still mathematically well-defined. $f$ admits a
generic Peter-Weyl decomposition $f(g) = \tr f_{I} D^I(g)$, where $f_I$ is a $d_I\times d_I$
matrix. The normalisation condition reads $\int f^2 = \sum_I \frac{1}{d_I}  \tr f_{I} \eps_I
f_{I}\eps_I=1$, with $\eps_I$ the isomorphism between the representation $V^I$ and its complex
conjugate as introduced previously. Then the kinetic term in the action would be expressed in term
of the convolution:
$$
f \circ f (g) = \sum_I \frac{1}{d_I} \tr f_{I} \eps_I f_{I} \eps_I D^I(g).
$$
Thus the kinetic term will depend explicitly on the momentum $P$ and not only on its norm
$P^2$.}

The interaction term involves a cubic and a quartic coupling. The cubic term is issued from quartic
interactions of the form $\int \vphi_f \psi \psi \psi$. The strength of the coupling is governed by
the coupling constant $\mu = \sqrt{6 \lambda} (\int f)$ which depends on the quartic coupling
$\lambda$.

This perturbed action is the action of an EFT with a non-linear redefinition of the momentum with
cubic and quartic interactions. Indeed, decomposing the central function $f$ on the basis of
characters, $f=\sum_I f_I \chi^I$ with the normalization constraint $\sum_I f_I^2=1$, we can
separate the kinetic part into a generalized momentum term, which vanishes on zero momenta (i.e. on
$g = 1\!\!1$), and a constant term yielding a massive contribution
\be
\forall g \in G, \;\;\; \mathcal{K}_f(g) = Q_f^2(g) - M_Q^2,
\ee
with the generalized momentum $Q_f$ given by
\be
Q_f^2(g) = \frac{\kappa^2}{2} \sum_{I \in \N} f_I^2 \left( 1- \frac{\chi^I(g)}{d_I} \right),
\ee
and the mass term by:
\be
M_Q^2 = \kappa^2 f_0^2 \, \leq \kappa^2.
\ee
Furthermore, since the (absolute value of the) character associated to the representation $I$ is
bounded by the associated dimension:~$\forall g \in G, \, \forall I \in \N, \, \mid
\frac{\chi^I(g)}{d_I} \mid \, \leq 1$, the squared momentum $Q^2_f$ is an
absolutely convergent series bounded from below by zero and from above by the squared Planck mass
\be
\forall g \in G, \;\;\;
0 \, \leq \, Q_f^2(g) \, \leq \, \kappa^2.
\ee
Accordingly, the kinetic term of the perturbed action is always positive which implies that the
theory is free from instabilities. Actually, the $\ka^2$ bound is loose, the true maximal bound
depends on the function $f$. The important point is that the momentum
$Q_f$ is still bounded, reflecting the compactness of the gauge group $\SU(2)$. Finally, we can
express the cubic coupling constant in terms of the mass $M_Q$. Since $M_Q=\ka\int f = \ka f_0$, it
is straightforward to obtain the relation\footnotemark:
$$
\lambda = \frac{\mu^2}{3!} \frac{\kappa^2}{M_Q^2},
$$
between the cubic and quartic coupling constants. Assuming that  $\lambda$ is held fixed, if $\int
f=0$, then both $M_Q$ and $\mu$ vanishes: the theory is massless and without cubic interaction. On
the other hand, assuming $\mu$ fixed, then the quartic coupling $\lambda$ is determined by its
ratio with the (dimensionless) mass $M_Q/\ka$: the cubic term will prevail for large $f_0$ while
$\lambda$ would blow up as $f_0$ goes to 0.

\footnotetext{This reminds of the relation between t'Hooft's coupling
constant $\lambda$ and the coupling constant of the Yang-Mills interaction $g_{\mbox{\tiny{YM}}}$:
$\lambda = g_{\mbox{\tiny{YM}}}^2 N$, where $N$ is the number of colors of the Yang-Mills theory.}

The generalized momentum $Q_f$ is in fact a non-linear redefinition of the momentum $P$ of the EFT.
As we have seen above, $Q_f$ is given by an infinite sum over representation labels $I$. Each term
of order $I$ generates powers of the momentum of degree $2I$. Indeed, the characters can be
expressed in terms of the Chebyshev polynomials of the second kind:
$$
\chi^I(\alpha) = U_{2I}(\cos \alpha)
=\sum_{r=0}^{I}(-1)^r \left( \begin{array}{c} 2I - r \\ r \end{array} \right)(2 \cos \alpha)^{2I -
2r}.
$$
This only involves powers of the squared cosine\footnotemark \,which can in turn be expressed as powers of
$\sin^2 \alpha = P^2(\alpha)/\kappa^2$. \footnotetext{This is the reason why we required that $f$ be a
function on $\SO(3)$ allowing only even modes in the Peter-Weyl decomposition, otherwise we would get
terms in $\sqrt{1 - (P /\ka)^2}$. Actually, to work with a well-defined EFT defined on $\SU(2)$ and
not only $\SO(3)$ requires a four-dimensional point of view \cite{duflo}. The four-momentum is
defined as: $\pi_i\equiv P_i$ and $\pi_4=\ka\sqrt{1 - (P /\ka)^2}$, with the mass-shell condition
$\pi^2=\ka^2$.} Accordingly, the characters are polynomials in $P^2$; $\chi^I = d_I -
\sum_{n=1}^{I} d_I c_{n}^{(I)} (\f1{\kappa} P)^{2n}$, and we obtain
\be
Q_f^2 \,=\, k_{1}[f] \, P^2 + k_{2}[f] \, \frac{P^4}{\kappa^{2}} + ... \, ,
\ee
where $k_{n}[f] = \frac{1}{2} \sum_I c_{n}^{(I)}{f_I^2} \in \R$. The explicit value of
these coefficients depends on the coefficients $c_{n} \in \R$ computable order by order through the
definition of the Chebyshev polynomials. For instance, the first order coefficient\footnotemark \,is
given by $k_{1}[f] = \f13 \sum_I f_I^2 \, C(I) =  \f13 \int f \Delta f$ with $C(I)=I(I+1)$ and
$\Delta$ respectively denoting the Casimir and Laplace operator on $\SU(2)$.

\footnotetext{We can compute these coefficient by matching the Taylor expansion of
the character in $\alpha$:
$$
\chi^I(\alpha)
= \chi^I (e^{2i \alpha J_z}) = \tr_I(1\!\!1) + i 2\alpha \tr_I J_z -2^2\frac{\alpha^2}{2!} \tr_I J_z^{2}
-i2^3\frac{\alpha^3}{3!} \tr_I J_z^{3}+ o(\alpha^4),
$$
with its expansion in $\sin^2\alpha$:
$$
\chi^I(\alpha) = U_{2I}(\cos \alpha) =
d_I - \sum_{n=1}^{I} d_I c_{n}^{(I)} \sin^{2n}\alpha= d_I - d_Ic_1^{(I)}
\alpha^2 + o(\alpha^4).
$$
Thus $d_Ic_1=2\tr_I J_z^{2}=\f23\tr_I \vec{J}^2=\f23d_II(I+1)$. Note that $\tr_I J_z=\tr_I
J_z^{3}=0$.
}

Note that $\sum_I f_I^2 \, C(I)$ does not necessarily converge for arbitrary $f\in L^2$. Actually
the higher order coefficients $k_n[f]$ will have similar expressions involving higher powers of
the Casimir $C(I)$. The simplest assumption in order to get a meaningful perturbative expansion is that the mode
decomposition of $f$ is finite, i.e involves a finite number of representations (this can be naturally achieved
by choosing $f$ appropriately or by replacing $\SU(2)$ with the quantum group $U_q(\mathfrak{su}(2))$ at root
of unity).

Hence, we have shown that the generalized momentum $Q_f$ can be defined perturbatively in inverse
powers of the Planck mass. Let us keep in mind that the Planck mass in $3d$ is simply the inverse
Newton constant $G_N$ and does not contain any factor in $\hbar$. Thus the perturbation of the
momentum $Q_f$ in $1/\ka$ is purely classical and does not require a quantum gravity interpretation.

The order zero, corresponding to the solution $f=0$, is simply a group field theory where the
matter degrees of freedom are frozen (trivial propagator). The first order gives the EFT dynamics
while the higher orders provide corrections to the EFT in $1/\ka$. The mass of the EFT and the
corrections are dictated by the function $f$, that is, by the classical solution \eqref{solution}
to Boulatov's theory. These high order modes correspond to higher derivatives in the matter action.
They can be interpreted as further gravitational corrections to the scalar field dynamics producing
new (unphysical) resonances. The question is now to understand how to interpret these new corrections,
keeping in mind that some specific solutions do not yield any corrective terms; they only appear in
the most general case. Our explanation of this fact is the following.

First, note that the presence of high order derivatives is a common fact
when working with effective field theories where the high energy modes have been integrated out.
For instance, the first order term in the expression of the generalized momentum, when $Q_f^2=P^2$,
already involves arbitrary high derivatives by virtue of the $\star$-product associated to the group
Fourier transform mapping the momentum group manifold onto non-commutative space-time \cite{pr3}.
However, here, we are generating corrections to a theory which is already deformed, not simply to flat QFT.
We interpret these corrections to the EFT's dynamics as a signature of the non-flatness of the metric
corresponding to the background geometry generated by the instanton around which we are perturbing.
Indeed, in the seminal work \cite{effqg,pr3}, the action for the EFT is derived from a theory of
{\itshape point particles} creating local conical defects in space-time. As a result, the EFT is formulated
on a flat (non-commutative) space-time and describes the matter field dynamics once that the
gravitational fluctuations around a flat metric have been integrated out. However, the
three-dimensional Einstein equations in presence of a scalar {\itshape field} allow, as classical
solutions, more complicated metrics than a simple locally flat, spinning cone metric generated by
point particles. Accordingly, it should possible to write the effective field theory of the scalar
field coupled to gravity on a curved geometry by integrating out the gravitational fluctuations
around the chosen background geometry. We interpret our instantonic solution as generating such a
background geometry solution to Einstein's equations in presence of a scalar field and not simply
in presence of a (finite) collection of point particles. From this perspective, we interpret the
deformed momentum $Q_f$ as the Fourier transform of a covariant derivative for a non-flat metric,
mapping the momentum group manifold to a curved space-time manifold. This momentum can then be
re-expressed as a non-linear function in the `flat' momentum $P$. This interpretation would be
confirmed by studying the field+gravity fluctuations around a non-trivial classical metric. We
postpone these investigations to future work.

We close this discussion with a remark concerning the generalized momentum $Q_f$. It is important to
note that the mass term $M_Q$ is the mass with respect to the generalized momentum $Q_f$. Namely, it is a
singularity of the propagator $(Q_f^2 - M_Q^2)^{-1}$. However, it is not the ``physical'' mass $M_P$
with respect to the flat momentum $P$ which would be defined as the singularity for the propagator
$(Q_f[P]^2 - M_Q^2)^{-1}$ expressed in term of $P$.


To conclude this section, we tune the classical solution \eqref{solution} as a mean to first obtain
exactly the EFT and then to compute the first order corrections.  Indeed, we can choose the
function $f$ such that its mode decomposition involves solely terms of order lower than two, i.e.,
$f_I = 0$, $\forall I \geq 2$. In this case, $f$  is a linear combination of the characters of the
trivial and adjoint representations:
$$
f = f_0 + f_1 \chi^1,
$$
with the constraint $f_0^2 + f_1^2=1$. This choice leads to the kinetic term
$$
\mathcal{K}_f(g) = k_{1}[f] \, P^2(g) - \kappa^2 f_0^2,
$$
that is, the action of a massive scalar field of mass $M_P^2= \frac{M_Q^2}{k_{1}[f]} = \f32
\kappa^2 \frac{f_0^2}{f_1^2}$. Note that $M_P$ is different from $M_Q$.

We can also generate massless actions simply by choosing an instanton associated to a function $f$
such that its mode decomposition involves a single spin $J$, i.e
$$
f = \pm \chi^J.
$$
Accordingly, the kinetic term of the perturbed action yields
$$
\mathcal{K}_f(g) =
P^2(g) \left[ \frac{1}{3} C(J) + \frac{1}{2} \sum_{n=1}^J c_{n}^{(I)} \left(
\frac{P(g)}{\kappa} \right)^{2 (n -1)} \right].
$$
We can readily see that, regarding the momentum $P$, the zero mass mode remains a solution but we
have generated new resonances as soon as $J$ is greater than $1$. These resonances will always be
unphysical, that is either complex or heavier than the Planck mass because of the positivity of
$Q_f$. For instance, we can compute the first correction by considering the background geometry
defined by the function $f =
\pm \chi^2$. This choice leads to a massless scalar field action with quartic momenta
$\mathcal{K}_f = 2 P^2 -
\frac{8}{5 \kappa^2} P^4$. Consequently, we obtain a new (unphysical) resonance $M_{(2)} = \kappa
\sqrt{5}/2 > \kappa$ in addition to the massless mode.

\section{Four dimensional outlook}

This last section is dedicated a discussion of the four dimensional extension of the above work.
The four dimensional GFT was first written by Ooguri in \cite{Ooguri}. The Feynmanology of the
theory generates spinfoam amplitudes of $4d$ BF theory with semi-simple, compact symmetry group $G$ discretized on
two-complexes dual to four dimensional triangulations. The action is a functional on the space of
complex fields on $G^{\times 4}$ given by
\bes
S_{4d}[\vphi] &=& \f12 \int_{G^{\times 4}} \vphi(g_1,g_2,g_3,g_4) \vphi(g_4,g_3,g_2,g_1) \\ \nn &&
- \frac{\lambda}{5!} \int_{G^{\times 10}} \prod_{i=1}^{10} dg_i
\vphi(g_1,g_2,g_3,g_4)
\vphi(g_4,g_5,g_6,g_7) \vphi(g_7,g_3,g_8,g_9) \vphi(g_9,g_6,g_2,g_{10}) \vphi(g_{10},g_8,g_5,g_1),
\label{4dgftaction}
\ees
where we require the field $\vphi$ to satisfy the same reality and symmetry requirements than in
two and three dimensions:
$$
\forall g\in G,\,\vphi(g_1g,g_2g,g_3g,g_4g)=\vphi(g_1,g_2,g_3,g_4), \qquad
\vphi(g_4,g_3,g_2,g_1)=\overline{\vphi}(g_1,g_2,g_3,g_4).
$$
The classical field equations are given by
\be
\vphi(g_4,g_3,g_2,g_1) =
\frac{\lambda}{4!} \int_{G^{\times 10}} \prod_{i=5}^{10} dg_i
\vphi(g_4,g_5,g_6,g_7) \vphi(g_7,g_3,g_8,g_9) \vphi(g_9,g_6,g_2,g_{10}) \vphi(g_{10},g_8,g_5,g_1).
\ee
An immediate generalization of the techniques developed above lead to the identification of the
following two-parameter family of classical solutions
\be
\vphi_{f_1,f_2}(g_1,g_2,g_3,g_4) = \sqrt[3]{\frac{4!}{\lambda}} \int_G \, dh \delta(g_1 h) f_1(g_2 h) f_2(g_3 h) \delta(g_4 h)
\ee
labeled by the coupled of functions $(f_1,f_2) \, \in \, L^2(G)^{\times 2}$ satisfying the
normalization constraint $\int dg\,f_1(g)f_2(g) =1$. An example of a such couple of functions is
given by the characters, $f_1=f_2=\chi_\rho$. This ansatz can easily be generalized to the group
field theories for the Barrett-Crane model \cite{BC} \footnotemark.

\footnotetext{
Following the original paper \cite{dfkr}, the Barrett-Crane spinfoam amplitude can be generated by
the same GFT action \Ref{4dgftaction} than the topological BF theory, but the field $\vphi$ must be
constrained. The symmetry group is the Lorentz group $G=\SO(\eta)$ where $\eta=(\sigma^2,+,+,+)$ with
$\sigma=1$ in the Riemannian case $\sigma=i$ in the Lorentzian case. We introduce two projectors,
$$
\PP\vphi(g_i)=\int_{\SO(\eta)} dg\, \vphi(g_ig),\quad
\QQ\vphi(g_i)=\int_{\SO(\bar{\eta})^4} [dh_i]^4 \,\vphi(g_ih_i),
$$
respectively projecting onto gauge invariant fields and $\SO(\bar{\eta})$-invariant fields, where
$\bar{\eta}$ is a three-dimensional flat metric whose isometry group leaves a given fixed internal
vector invariant. Accordingly, its signature fixes $\SO(\bar{\eta})$ to be $\SO(3)$ in the
Riemannian case and $\SO(3)$ or $\SO(1,2)$ in the Lorentzian case. The Barrett-Crane model is
defined by the restriction $\vphi\in\,{\rm Im}\PP\QQ$. Since $\PP$ and $\QQ$ do not commute, the
operator $\PP\QQ$ is not a projector. This creates normalisation ambiguities for the field $\vphi$
which lead to ambiguities in the precise GFT interaction for the constrained model
\cite{dan_gftcoupling}. We can generalize the classical solutions found for the full GFT to this
constrained GFT by applying the operator $\PP\QQ$ to the field
$\delta(g_1)f_1(g_2)f_2(g_3)\delta(g_4)$. In particular, the $\QQ$ projection implies that the
fields will decompose only onto the simple representations of $\SO(\eta)$ and that we will work
with the spherical kernels on $\SO(\eta)$ instead of the characters.}

Followed the proposal developed in this paper, we can try perturbing the $4d$ GFT around these classical
solutions by a lower dimensional phase and analyze the resulting dynamics. It is straightforward to
check that two-dimensional perturbations do not acquire any non-trivial propagator. Then we would need to move
to $3d$ perturbations. We expect that the obtained action will lead to a field theory whose
fundamental excitations are no longer particles but one-dimensional strings, i.e., a string field
theory.

In such a framework, the Feynman diagrams of the theory are networks $\Sigma$ of elementary
surfaces; the propagators are assigned to surfaces along which two three-cells (the Feynman loops
of the theory) are glued together. Accordingly, discretizing the surface $\Sigma$ and choosing a
triangulation $\T$ adapted to such diagrams ($\Sigma \subset \T$), it is possible to perform a
(four dimensional) duality transformation on the associated amplitudes by replacing the `momenta'
associated to the triangles $\Delta \subset \T$ by dual variables assigned to the tetrahedra
sharing the given triangle, or equivalently to the dual edges $e^*
\subset \T^*$ of the dual triangulation. We expect that these transformed amplitudes correspond to
the spinfoam model \cite{AW} of the canonical Baez-Perez proposal \cite{BP}:
\be
\mathcal{Z} (\T,\Sigma) =
\sum_{\{\rho_{\Delta}\}} \prod_{\Delta \notin \Sigma}
d_{\rho_{\Delta}} \prod_{\Delta \in \Sigma} \mathcal{P}_T(\rho_{\Delta}) \prod_{s} \{ 15 j\}_s,
\ee
where the symbol $s$ labels the four-simplices of the triangulation, $\{ 15 j\}_s$ denotes a
$15j$ symbol constructed from the ten representations and the five intertwining operators
associated to the ten triangles and five tetrahedra building a given four-simplex $s$,
$\mathcal{P}_T(\rho_{\Delta})$ is the string `propagation' term depending on the unitary,
irreducible representation label $\rho_{\Delta}$ of $G$ associated to the triangle $\Delta$ and on
the string tension $T$. The details of this four dimensional construction are currently under study
\cite{EW}.

\section{Conclusion}

The study of the EFT of matter coupled to quantum gravity achieved in this paper has revealed two
new aspects. The first is that this field theory can be reexpressed as a generalized matrix model.
The non-trivial dynamics imply that the matrices can change size during propagation. The novelty is
the interpretation of the dimensions of the matrices from the three-dimensional quantum gravity
perspective. It appears clearly that the size of the matrix fields encodes the physical length
quantum numbers of the Ponzano-Regge model. The second major aspect unraveled by the present work
is the role of classical solutions to the GFT. We have identified a one parameter family of
solutions to Boulatov's field equations such that perturbations around these non-trivial
geometrical backgrounds generate the dynamics of the EFT. In fact, the most general solutions lead
to higher order derivative corrections to the EFT. We have interpreted these corrections as the signature of the
non-flatness of the background geometry around which we have integrated the gravitational fluctuations.
This geometry is generated by the instantonic solutions to the GFT, and can be understood as the geometry
associated to a solution to the Einstein equations in presence of a scalar field. As a result, the associated
EFT would be defined on a curved space-time. This is in contrast with
the original setting of the (flat) EFT describing the dynamics of matter once that the gravitational fluctuations
around a flat background geometry, punctured with local topological defects, have been integrated out.
The proof of these interpretations will be studied elsewhere. To make progress in this direction, it is now urgent
to study in great detail the precise relationship between the group field theory and general relativity as classical
theories.

The lessons of this paper are a two-fold. First, the crucial role played by the non-perturbative
aspects of the GFT. It has been suspected for long that these effects should play a significant
role in the formulation of quantum gravity theories \cite{gftlaurent,thesis}. Here, for the first
time, we have explicitly shown that non-trivial solutions to the classical field equations could
generate non-trivial geometrical properties of spacetime like for instance matter propagation on a
quantum gravity background. This leads to the
second point emphasized by our work: matter is defined by excited geometry states.
Indeed, we have generated the dynamics of matter purely from
field solutions to the (vacuum) GFT. We have not introduced matter degrees of freedom by hand, as
in the pure spinfoam context; matter appears as a particular phase of the field theory of
simplicial geometry, i.e., the GFT. In this sense, the work presented here is in striking contrast
with the GFT models containing extra data to model the particle content of the theory
\cite{gftwithparticles,kirillGFT}. In fact, our work shows that no extra data is needed: {\it
matter is a particular phase of the geometry} and is already somehow contained in our quantum
gravity models. It would be interesting to extend our results to the non-scalar case, that is, generate
the effective field theory of spinning fields from the GFT following the procedure developed here.

It seems therefore tempting to apply the same type of rationale to the four-dimensional case. We
have discussed the prototype ideas in the last section of this article. It now appears clearly that
the extension of the three-dimensional spinfoam quantum gravity formalism to $4d$ will require the
introduction of string-like excitations of the geometry. We have derived solutions to Ooguri's
field equations and the next logical step is to study the perturbations around the background
defined by such a solution. Of course, four-dimensional BF theory is non-geometrical until the
simplicity constraints are implemented. Accordingly, the physical situation will require some extra
incomes, like for instance starting from the Barrett-Crane GFT \cite{dfkr}.


\section*{Acknowledgements}

We thank Laurent Freidel for many conversations on the relevance of classical solutions to the GFT in the
study of spinfoam models and Daniele Oriti for general discussions on the subject.

\appendix

\section{Field Solutions to the Effective Non-Commutative QFT}

In this Appendix, we show how to compute some classical solutions to the equations of
motion of the effective non-commutative QFT defined by the action $S_{eff}$. On the one hand, it is
interesting to compare them to the classical solutions of the 2d group field theory and see the
effect of the new kinetic term with $P^2(g)$. On the other hand, it provides us with new classical
solutions to the Boulatov-Ooguri 3d group field theory. Determining the classical solutions of the
GFT will naturally provide informations on the semi-classical regime of spinfoam models
\cite{gftlaurent,inprep}.

We work with a cubic interaction term, but we hope that these considerations will be generalizable
to higher order interactions. The equation of motion of the 2d GFT is \cite{thesis}:
\be
\phi(g)=\f\lambda2\int_G dh\, \phi(h)\phi(h^{-1}g) =\f\lambda2\phi\circ\phi (g),
\ee
where $\circ$ stands for the convolution product on $\SU(2)$. The classical field $\phi$ is thus a
projector (up to the factor $\lambda/2$ for $\circ$. Looking for solutions invariant under
conjugation\footnotemark i.e which only depends on the angle of rotation $\theta$ of the group
element $g$, the only solutions are actually the characters $\chi^j$:
$\phi(g)=d_j\chi^j(g)\lambda/2$ for each spin $j\in\N/2$ provides an infinite number of solutions
of the equation of motion.

Now let us look at the equation of motion for the effective QFT for a massless scalar field. The
only difference is the factor $P^2(g)$:
\be
P^2(g)\phi(g) =\f\lambda2\phi\circ\phi (g).
\ee
Since $P^2(g)$ contains $\chi^1(g)$, these equations is going to couple the different
representations and the solutions will not be as simple as for the 2d GFT. In order to solve this
equation, it is more convenient to expand the field $\phi$ in representations. Once again, we will
only look for fields invariant under conjugation. We will also assume that $\phi$ is even (it is a
field over $\SO(3)$) and only decomposes onto integer spins. Then we decompose $\phi$ onto the
characters: $\phi(g)=\sum_{j\in\N}
\phi_j\chi_j$. Taking into account that $\chi^1\chi^j=(\chi^{j-1}+\chi^j+\chi^{j+1})$ for $j\ge 1$,
the equation of motion becomes:
$$
\f34\sum_j\phi_j\chi^j -\f14\phi_0\chi^1 -\f14\sum_{j\ge 1} \phi_j(\chi^{j-1}+\chi^j+\chi^{j+1})
\,=\,
\f\lambda2\sum_j \f{\phi_j^2}{d_j}\chi_j.
$$
This translates to a set of recursion relations on the $\phi_j$'s:
\be
\left|\begin{array}{lcl}
3\phi_0-\phi_1 &=&\lambda\phi_0^2,\\
\phi_j-\f12(\phi_{j-1}+\phi_{j+1})&=&\lambda \f{\phi_j^2}{d_j}.
\end{array}\right.
\label{recur}
\ee
The first equation gives $\phi_1$ in term of the initial value $\phi_0$. The other equation
determines $\phi_{j+1}$ in term of $\phi_j$ and $\phi_{j-1}$ as soon as $j\ge 1$. We see that we
have fewer solutions as above and that the structure of the solutions are actually very different
due to the coupling between representations induced by the factor $P^2(g)$. We have not been able
to obtain a closed form for the classical field $\phi(g)$ in term of the normalisation $\phi_0$. We
can nevertheless discuss the asymptotical behavior of $\phi_j$ when $j$ goes to $\infty$. Indeed
recognizing the left hand term in \Ref{recur} as a discretized second derivative\footnotemark, we
see that $\phi_j$ should go asymptotically as $1/j$. More precisely, the asymptotical solution is
$-2/\lambda j$.

\footnotetext{
$\lambda=0$ is a special case. The interaction term $\phi_j^2$ disappear from the equations of
motion and we are dealing with a free scalar field. It is then straightforward to check that we
have a two-parameter family of solutions of the type $\phi_j=aj+b, j\ge 1$, which have a vanishing
discretized second derivative. As soon as $\lambda$ is turned on, the physical content of the
theory changes completely.}

This is actually be compared to the classical equation in the continuum:
$$
\Delta \phi = -\lambda \phi^2.
$$
Assuming $\phi$ to be invariant under rotation and thus to only depend on the radial coordinate
$r$, this equation becomes:
$$
\pp_r^2(r\phi)\,=\,-\lambda \f{(r\phi)^2}{r}.
$$
This equation looks like the continuum limit of the recursion relation \Ref{recur}. Its obvious
solution is $(r\phi(r))=-2/\lambda r$. To make the correspondence more explicit, we would like to
identify $r\phi(r)$ to $\phi_j$ and the discrete difference $(\phi_{j-1}+\phi_{j+1})/2-\phi_j$ to
the second derivative $\pp_r^2(r\phi)$. This is actually realized through the Duflo map introduced
in \cite{duflo}, where the authors prove that this correspondence is made true in the context of the
non-commutative geometry and $\star$-product underlying the effective theory $S_{eff}$ for matter
coupled to 3d quantum gravity.

\section{Full Perturbations of the 3d GFT}

We now describe the full perturbed action around the considered classical solutions
to the 3d group field theory without restricting ourselves to the 2d sector, that is we explicitly
compute the full action $S_{3d}^f[\vphi]\,\equiv\,S_{3d}[\vphi_f+\vphi]-S_{3d}[\vphi_f]$ defined in
eqn.\Ref{Sf}. It is straightforward to compute:
\bes
S^f_{3d}[\vphi]\,=\, S_{3d}[\vphi] &-&\f12\int_{G^{\times 3}} \prod_{i=1}^3 dg_i \,\vphi(g_1,g_2,g_3)\vphi(g_3,g_2,g_1)\int_G
dg\,f(g)f(gg_1g_3^{-1}) \nn\\ &-&\int_{G^{\times 4}} \prod_{i=1}^4 dg_i
\,f(g_2g_1^{-1})\vphi(g_1,g_2,g_3)f(g_4g_1^{-1})\vphi(g_3,g_4,g_1) \nn\\
&-&\sqrt{\f{\lambda}{3!}}\int_{G^{\times 5}} \prod_{i=1}^5 dg_if(g_4g_1^{-1})\vphi(g_1,g_2,g_3)\vphi(g_3,g_4,g_5)\vphi(g_5,g_2,g_1).
\ees
We obtain two new quadratic terms which produce a non-trivial propagator for the GFT and a new
cubic interaction term with coupling constant $\sqrt{\lambda}$. Here, we insist on the fact that we
are simply perturbing the GFT action around a non-trivial field configuration. Therefore the
non-perturbative partition function does not change at all although the structure of its
perturbative expansion might get modified.

The field $\phi(g_1,g_2,g_3)$ is gauge-invariant. Thus it is actually a function of two ``loop
variables", $g_1g_3^{-1}$ and $g_2g_3^{-1}$. Following the logic of the paper, we introduce the
following generic multi-component ansatz for the field:
\be
\vphi(g_1,g_2,g_3)=\sum_\alpha \psi_\alpha(g_1g_3^{-1})A_\alpha(g_2g_3^{-1}),
\ee
where $\alpha$ is an abstract index. We consider the $\psi_\alpha(g)$ as field variables while we
hold the $A_\alpha(g)$ fixed. However this is only a point of view and both $\psi_\alpha$ and
$A_\alpha$ could be considered as variables. Assuming that the field $\vphi$ is still real, the
kinetic term of the action $S^f_{3d}$ with a non-trivial background field now reads:
\be
\f12\int_G dg\, \psi_\alpha(g)\overline{\psi}_\beta(g)\,\left[\left(1-\int_G dh\, f(h)f(hg)\right)\,\int_G dk
A_\alpha(k)\overline{A}_\beta(k)
\,-\,\left(\int_G dh\, f(h)A_\alpha(hg)\right)\left(\int_G dk\,
f(k)\overline{A}_\beta(kg)\right)\right].
\ee
This allows a coupling between the abstract (internal) indices $\alpha,\beta$ labeling the field
components and the momentum $g$. This might allow to derive actions for matter fields with spin.



\begin{thebibliography}{10}

\bibitem{effqg}
  L.~Freidel and E.~R.~Livine,
  {\it Effective 3d quantum gravity and non-commutative quantum field theory},
  Phys.\ Rev.\ Lett.\  {\bf 96}, 221301 (2006),
  [arXiv:hep-th/0512113]

\bibitem{SF}
A. Perez, {\it Spin foam models for quantum gravity}, Class. Quant. Grav.
{\bf 20} R43 (2003), [arXiv:gr-qc/0301113];

D. Oriti, {\it Spacetime geometry from algebra: Spin foam models for non-perturbative  quantum
gravity},  Rept. Prog. Phys. {\bf 64} 1489-1544 (2001), [arXiv:gr-qc/0106091];

J. C. Baez, {\it An introduction to spin foam models of BF theory and quantum gravity}, Lect. Notes
Phys. {\bf 543} 25-94 (2000), [arXiv:gr-qc/9905087]






\bibitem{BC}
J. W. Barrett and L. Crane, {\it Relativistic spin networks and quantum gravity}, J. Math. Phys.
{\bf 39} 3296-3302 (1998), [arXiv:gr-qc/9709028]; J. W. Barrett and L. Crane, {\it A Lorentzian
signature model for quantum general relativity'}, Class. Quant. Grav. {\bf 17} 3101-3118 (2000),
[arXiv:gr-qc/9904025]

\bibitem{BCd}
L. Freidel, K. Krasnov and R. Puzio, {\it BF description of higher-dimensional gravity theories},
Adv. Theor. Math. Phys. {\bf 3} 1289-1324  (1999), [arXiv:hep-th/9901069]

\bibitem{rr}
  M.~P.~Reisenberger and C.~Rovelli,
  {\it Spacetime as a Feynman diagram: The connection formulation},
  Class.\ Quant.\ Grav.\  {\bf 18} 121 (2001),
  [arXiv:gr-qc/0002095];
  {\it Spin foams as Feynman diagrams},
  [arXiv:gr-qc/0002083].

\bibitem{SFrep}
A. Perez, ``The spin-foam-representation of loop quantum gravity'', [arXiv:gr-qc/0601095]

\bibitem{gftlaurent}
  L.~Freidel,
  {\it Group field theory: An overview},
  Int.\ J.\ Theor.\ Phys.\  {\bf 44} 1769 (2005),
  [arXiv:hep-th/0505016]

\bibitem{dan_review}
  D.~Oriti,
  {\it The group field theory approach to quantum gravity},
  [arXiv:gr-qc/0607032]

\bibitem{2dsf}
E.~R.~Livine, A.~Perez and C.~Rovelli,
{\it 2D manifold-independent spinfoam theory},
Class.\ Quant.\ Grav.\  {\bf 20} 4425 (2003), [arXiv:gr-qc/0102051]

\bibitem{PR}
G. Ponzano and T. Regge, ``Semiclassical limit of Racah coefficients'', {\em Spectroscopic and
group theoretical methods in physics, North-Holland Publ.}, 1968

\bibitem{3dsf}
  D.~V.~Boulatov,
  {\it A Model of three-dimensional lattice gravity},
  Mod.\ Phys.\ Lett.\ A {\bf 7} 1629 (1992),
  [arXiv:hep-th/9202074]

\bibitem{dfkr}
  R.~De Pietri, L.~Freidel, K.~Krasnov and C.~Rovelli,
   {\it Barrett-Crane model from a Boulatov-Ooguri field theory over a  homogeneous
  space},
  Nucl.\ Phys.\ B {\bf 574} 785 (2000),
  [arXiv:hep-th/9907154]

\bibitem{laurent&david}
  L.~Freidel and D.~Louapre,
  {\it Non-perturbative summation over 3D discrete topologies},
  Phys.\ Rev.\ D {\bf 68} 104004 (2003),
  [arXiv:hep-th/0211026]

\bibitem{pr1}
  L.~Freidel and D.~Louapre,
   {\it Ponzano-Regge model revisited. I: Gauge fixing, observables and interacting spinning particles},
  Class.\ Quant.\ Grav.\  {\bf 21}, 5685 (2004)
  [arXiv:hep-th/0401076]

  \bibitem{aristide}
A. Baratin and L. Freidel, {\it Hidden Quantum Gravity in 3-D Feynman diagrams},
[arXiv:gr-qc/0604016]; A. Baratin, L. Freidel, {\it Hidden Quantum Gravity in 4-D Feynman diagrams:
Emergence of spin foams}, [arXiv:hep-th/0611042]

\bibitem{barrett}
J. W. Barrett, {\it Feynman loops and three-dimensional quantum gravity}, Mod.Phys.Lett. {\bf A20}
1271 (2005), [arXiv:gr-qc/0412107]; J. W. Barrett, {\it Feynman diagrams coupled to
three-dimensional quantum gravity}, Class.Quant.Grav. {\bf 23} 137-142 (2006),
[arXiv:gr-qc/0502048]

\bibitem{pr3}
   L.~Freidel and E.~R.~Livine,
   {\it Ponzano-Regge model revisited III: Feynman diagrams and effective  field
  theory},
  Class.\ Quant.\ Grav.\  {\bf 23}  2021 (2006),
  [arXiv:hep-th/0502106]



\bibitem{gftwithparticles}
  L.~Freidel, D.~Oriti and J.~Ryan,
  {\it A group field theory for 3d quantum gravity coupled to a scalar field},
  [arXiv:gr-qc/0506067];
  D.~Oriti and J.~Ryan,
  {\it Group field theory formulation of 3d quantum gravity coupled to matter fields},
  [arXiv:gr-qc/0602010]

\bibitem{kirillGFT}
   K.~Krasnov,
  {\it Quantum gravity with matter via group field theory},
  [arXiv:hep-th/0505174]

\bibitem{thesis}
  E.R.~Livine,
  {\it Loop gravity and spin foam: Covariant methods for the non-perturbative
  quantization of general relativity}, PhD thesis 2003 - Centre de Physique Th\'eorique (Marseille,
  France),  [arXiv:gr-qc/0309028]

\bibitem{inprep}
L.~Freidel and  E.~R.~Livine,
  {\it Non-perturbative structures for Spinfoams: Instantons for the Group Field Theory},
  in preparation

\bibitem{duflo}
  L.~Freidel and S.~Majid,
   {\it Noncommutative harmonic analysis, sampling theory and the Duflo map in 2+1 quantum gravity},
  [arXiv:hep-th/0601004].

\bibitem{Nexpansion}
G. 'tHooft, {\it A Planar Diagram Theory for Strong Interactions}, Nucl.Phys. {\bf B72} 461 (1974); \\
E. Witten, {\it Baryons In The 1/N Expansion}, Nucl. Phys. {\bf B160} 57 (1979)



\bibitem{Ooguri}
H. Ooguri, {\it  Topological lattice models in four-dimensions},  Mod.Phys.Lett. {\bf A7} 2799-2810
(1992), [arXiv:hep-th/9205090]

\bibitem{dan_gftcoupling}
D. Oriti,
{\it Boundary terms in the Barrett-Crane spin foam model and consistent gluing},
Phys.Lett. B532 (2002) 363-372, [arXiv:gr-qc/0201077]

\bibitem{AW}
W. J. Fairbairn and A. Perez, {\it Quantisation of string-like sources coupled to BF theory :
Physical scalar product and topological invariance}, in preparation

\bibitem{BP}
J. C. Baez and A. Perez, {\it Quantization of strings and branes coupled to BF theory},
[arXiv:gr-qc/0605087]

\bibitem{EW}
W. J. Fairbairn and E. R. Livine, {\it String theory as a phase of the four dimensional group field
theory}, in preparation

\end{thebibliography}
\end{document}